 \documentstyle[12pt]{article}
 \def \bfgr #1{ \mbox {{\boldmath $#1$}}}
\begin{document}
\begin{center}
{\Large \bf Relativistic Calculation of\\[3mm]
Structure Functions $b_{1,2}(x)$ of the Deuteron.}

\vskip 5mm 

{A.Yu.  Umnikov}
\footnote{INFN Postdoctoral Fellow}

 \vskip 5mm 

{\em Department of Physics, University of Perugia, and
INFN, Sezione di Perugia,\\
via A. Pascoli, Perugia, I-06100, Italy.}

\end{center}
\vskip 5mm

\begin{abstract}
The  structure functions $b_{1,2}^D(x)$ of the deuteron 
are studied within covariant approach. 
It is shown that usual nonrelativistic
convolution model   result in incorrect
behavior of this  structure functions at small $x$
and violates the exact sum rules. Realistic calculations 
are carried out with the Bethe-Salpeter amplitude of the deuteron
and compared with the nonrelativistic results.
\end{abstract}


\section{Introduction}

The study of the deep inelastic lepton scattering  
with polarized 
targets and beams provides   refined information about the
quark content of   hadrons. These days much attention is attracted
to the nucleon's
spin-dependent  Structure Functions (SF), $g_{1,2}^N$
 (see e.g. review~\cite{efremov} and references therein). 
The SFs, $g_{1,2}$, is
 the simplest example of the spin-dependent
SFs, which exist for all targets with
a nonzero spin, starting $s=1/2$. 
At the same time, hadrons with a spin higher than $1/2$ have
additional spin-dependent SFs~\cite{jm}.

The spin-dependent SFs
$b_{1,2}(x)$\footnote{The notation $b_{1,2}$  is used, following ref.~\cite{hjm}.}
 of  spin-1 hadrons
have been studied on a few occasions,
including the vector mesons and deuteron~\cite{et,fs1,hjm,ku}. 
For    mesons some qualitative estimates have 
been done, but ``a real understanding of $b_1(x)$ at the quark level is not 
yet available''~\cite{hjm}.
The only exact sum rules for  mesons  
   have been proposed
 by Efremov and Teryaev~\cite{et}:
\begin{eqnarray}
\int\limits_0^1 b_1(x) dx  =0.
\label{sr1}\\
  \int\limits_0^1 b_2(x) dx =0.
\label{sr2}
\end{eqnarray}

It would be unrealistic to suggest that these sum rules 
will be experimentally verified for the mesons  any time soon.
However, they are
 independent of the target, i.e. it is supposed to be
valid for the spin-1 nuclei as well. 
The deuteron
is the most probable candidate
for its SFs $b_{1,2}$  to be measured.
Indeed, a number of 
 deep inelastic experiments
on the polarized deuterons  are run  or being prepared
at the world's best facilities,
such as SLAC, CERN, CEBAF and DESY. These experiments are usually
aiming to extract a
neutron  SF $g_1$, however, in principle,
{ the same data} can be used to obtain  SFs
$b_{1,2}$. The problem is caused only by the yet insufficient
accuracy of measurements. 
Realistic calculations with
the deuteron wave functions in the Bonn and Paris potentials
have been done in ref.~\cite{ku,ukk}. 
The preliminary results within the Bethe-Salpeter formalism
have also been  presented~\cite{ukk}. It was shown that 
relativistic calculations differ from nonrelativistic
ones, but the reasons have not been discussed. 
The sum rules (\ref{sr1})-(\ref{sr2}) 
have not been analyzed in these papers but the displayed shape of the
 SFs makes questionable the validity of the sum rules.

This letter presents a study of the deuteron  SFs
$b_{1,2}^D(x)$ within a covariant approach, based
on the relativistic convolution formalism
for the deep inelastic scattering~\cite{fs2,land,km,msm}
and the Bethe-Salpeter formalism for the deu\-te\-ron
bound state~\cite{tjond,uk,ukkk}. 
The issue of the
sum rules is specially addressed. It is shown that the nonrelativistic
convolution model is not relevant to calculate  
 SFs $b_{1,2}^D(x)$  at small $x$, because
this model  {\em inevitably} 
 breaks at least one of the sum rules (\ref{sr1}) or (\ref{sr2}) .
At the same time, the relativistic calculations
 based on the Bethe-Salpeter
amplitude for the deuteron are {\em consistent}
with both of the sum rules. The  SFs,
$b_{1,2}^D(x)$, are calculated within both
relativistic and nonrelativistic approaches.

\section{ Structure functions in the relativistic impulse approximation}

The nucleon contribution to the deep inelastic 
scattering of  electrons  off the deuteron is defined by the triangle
diagram (see Fig.~\ref{tre})~\cite{fs2,land,km,msm}:
\begin{eqnarray}
\langle \hat W \rangle_M = i\int \frac{d^4p}{(2\pi)^4}{\sf Tr}\left\{
\bar\Psi_M(p_0,{\bfgr p})\hat W(q,p_1) \Psi_M(p_0,{\bfgr p}) (\hat p_2-m)
\right \},
 \label{me}
\end{eqnarray}
where $p_{1,2} = P_D/2\pm p= (M_D/2 \pm p_0, \pm {\bfgr p})$, $P_D$
 is the deuteron momentum, $M_D$ is the deuteron mass,
 $p = (p_0,{\bfgr p})$
is the relative momentum of nucleons,
$\hat W$ is an operator appropriate 
for the process (see e.g. and references therein~\cite{msm,mst,kmt}) and 
 $\Psi_M(p_0,{\bfgr p})$ is the Bethe-Salpeter amplitude for the deuteron with
$M$ being  the deuteron's total momentum projection. 
 The original
convolution model for  deep inelastic scattering 
is reproduced by choosing operator
$\hat W$ in the form:
\begin{eqnarray}
\hat W = \frac{\hat q}{2pq}W^N_{\mu\nu}(q,p_1),
 \label{op}
\end{eqnarray}
where $m$ is  the nucleon mass, $q=(\nu,0,0,-\sqrt{\nu^2+Q^2})$
 is the momentum transfer,
  $ Q^2=-q^2$ and $W_{\mu\nu}(p,q)^N$
is the hadron  tensor of nucleon: 
\begin{eqnarray}
&&  W_{\mu\nu}^N(q,p) =
 \label{hten} \\
&& \left ( -g_{\mu\nu} +\frac{q_\mu q_\nu}{q^2}\right ) F_1^N(x,Q^2)
 + \left ( p_\mu - q_\mu \frac{pq}{q^2}  \right ) 
  \left ( p_\nu - q_\nu \frac{pq}{q^2}  \right ) \frac{F_2^N(x,Q^2)}{pq},
  \nonumber
\end{eqnarray}
where $x = Q^2/(2pq)$ anf $F_{1,2}^N$ 
are the nucleon  SFs.
 The small effects of the
off-mass-shell deformation
of the nucleon tensor\cite{mst,kmt,uoff} are not considered in the present letter,
because these effects do not affect sum rules (\ref{sr1})-(\ref{sr2})  and do not noticeably
change the absolute values of the  SFs. That is why 
   SFs $F_{1,2}^N$ in (\ref{hten}) do not depend on $p^2$, but 
  $q^2$ and $pq$.

One can extract   scalar    SFs from the hadron of
the deuteron. For instance, for $F_2$ one gets:
\begin{eqnarray}
F_2^D(x_N,Q^2,M) &=& \frac{i}{2M_D}\int \frac{d^4p}{(2\pi)^4}
F_2^N \left( \frac{x_N m}{p_{10}+p_{13}}, Q^2\right)\nonumber \\[2mm]
&&\quad\quad {\sf Tr}\left\{
\bar\Psi_M(p_0,{\bfgr p})(\gamma_0+\gamma_3) \Psi_M(p_0,{\bfgr p}) (\hat p_2-m)
\right \},
 \label{f2m}
\end{eqnarray}
where $x_N=Q^2/(2m\nu)$ is the Bjorken scaling 
variable\footnote{Note that the ``native'' deuteron variable
is $x_D = (m/M_D)x_N$, however
 $x_N$ is used more often.}, 
i.e. this is $x$ 
for the on-mass-shell nucleon at rest,
 $p_{10}$ and $p_{13}$ are the time and 3-rd components of the
1-st nucleon momentum. 
Formula (\ref{f2m}) has not been averaged over the projection of the deuteron 
total momentum, $M$, since in the present form it gives an understanding what
the  SF $b_2^D$ is. Indeed, eq.~(\ref{f2m}) gives two
independent `` SFs'', with $M = \pm 1$ and $M=0$, which are
related to the usual spin-independent  SF, $F_2^D$, and
a new  SF, $b_2^D$:
\begin{eqnarray}
&&F_2^D(x_N,Q^2) = \frac{1}{3} \sum_{M=0,\pm 1} F_2^D(x_N,Q^2,M),
 \label{f2}\\[1mm]
&& b_2(x_N,Q^2) = F_2^D(x,Q^2,M=+1)-F_2^D(x,Q^2,M=0),\label{b2}\\[3mm]
&& F_2^D(x_N,Q^2,M=+1) = F_2^D(x_N,Q^2,M=-1).\label{pm}
\end{eqnarray}
Of course, instead of (\ref{f2}) and (\ref{b2}) any other linearly 
independent combination of the functions $F_2^D(x,Q^2,M)$ can be chosen.

Note,  SFs $F_2^D(x,Q^2,M)$ are independent of the 
lepton polarization, therefore, both  SFs, $F_2^D$ 
and $b_2^D$, can be measured in  experiments with an unpolarized
lepton beam and polarized deuteron target. In view of  eq.~(\ref{pm}),
only one  of the  SFs $F_2^D(x,Q^2,M)$ is needed, in addition
to the spin-independent $F_2^D(x,Q^2)$, in order to obtain
$b_2(x,Q^2)$. The other  SF, $b_1^D$, is related
to the deuteron  SF $F_1^D$, the same way as $b_2^D$ is
related to $F_2^D$, viz. via eqs.~(\ref{f2}), and $b_2^D = 2xb_1^D$.

\section{ Singularities of the triangle diagram and sum rules}

It has been previously shown~\cite{fs2,land,km,msm} 
how a singular structure of the
triangle graph (Fig.~\ref{tre}) rules the behavior of the
spin-independent   SF $F_2^D$. In particular,
it has been found that the relativistic impulse approximation
satisfies the unitarity and provides the correct kinematical
region
of the variable $x_N$. However,  both 
these properties of the exact covariant amplitude are
broken in  practical 
calculations, when  nonrelativistic wave functions of the deuteron 
are used. In this case  one can refer to the argument that such
deviations are small, and  are not important
for phenomenology.
At the same time, a realistic Bethe-Salpeter 
amplitude of the deuteron serves ideally for a 
consistent phenomenological application of the covariant theory
of the processes on the bound nucleons. 

In order to calculate  SFs, (\ref{f2m})-(\ref{b2})
 and analyze the sum rules, the singularities of the
 triangle diagram should be explicitly taken into account.
 To do that, eq.~(\ref{f2m}) is rewritten as:
\begin{eqnarray}
&&F_2^D(x_N,Q^2,M) =\nonumber\\
&&\quad \frac{i}{2M_D}\int \frac{d^4p}{(2\pi)^4} 
F_2^N \left( \frac{x_N m}{p_{10}+p_{13}}, Q^2\right)
\frac{1}{(p_1^2-m^2+i\epsilon)^2(p_2^2-m^2+i\epsilon)}\nonumber \\
&&\quad
{\sf Tr}\left\{
\bar\phi_M(p_0,{\bfgr p})
 (\hat p_1+m)(\gamma_0+\gamma_3) (\hat p_1+m)
 \phi_M(p_0,{\bfgr p}) (\hat p_2+m)
\right \},
 \label{sing1}
\end{eqnarray}
where $\phi_M(p_0,{\bfgr p}) = (\hat p_1-m)\Psi_M(p_0,{\bfgr p})(\hat p_2-m)$
is the Bethe-Salpeter vertex functions of the deuteron.

 Analysis of singularities in the complex $p_{2+}$-plane
  allows for one  analytical 
   integration in (\ref{sing1})~\cite{km}. Being translated into 
 variables which are used in the present paper, this integration
 is equivalent to picking  the residue in the second nucleon pole,
 $p_{20} = \omega  = \sqrt{m^2+{\bfgr p}^2}$ or 
 $p_0 = M_D/2 - \omega $,
 in the complex $p_{0}$-plane when both of 
 the following conditions are satisfied:
\begin{eqnarray}
0 < \omega  - p_{3} < M_D.
 \label{sing2}
\end{eqnarray}
The contribution of the region of $p$  beyond (\ref{sing2}),
 into integral (\ref{sing1}), is zero, i.e. different poles cancel each other. 
 Note,  that in the required pole
  $p_{10} = M_D -
 \omega $. Calculating residue in (\ref{sing1}), one gets: 
  \begin{eqnarray}
&&F_2^D(x_N,Q^2,M) = \frac{1}{2M_D}\int \frac{d^3{\bfgr p}}{(2\pi)^3} 
F_2^N \left( \frac{x_N m}{M_D- \omega  +p_{3}}, Q^2\right)
\label{sing3}\\
&&\quad\quad\quad\quad\quad\quad\quad\quad \Theta(M_D-\omega  + p_{3} )
\frac{1}{2\omega M_D^2 (M_D-2\omega  )^2}
\nonumber \\
&& 
{\sf Tr}\left\{
\bar\phi_M(p_0,{\bfgr p})
 (\hat p_1+m)(\gamma_0+\gamma_3) (\hat p_1+m)
 \phi_M(p_0,{\bfgr p}) (\hat p_2+m)
\right \}_{p_0 = \frac{M_D}{2}-\omega },
\nonumber
\end{eqnarray}
where the $\Theta$-function guaranties the right of  conditions (\ref{sing2}),
the left condition is always satisfied.
 
It is useful to rewrite (\ref{sing3}) in the convolution form:
\begin{eqnarray}
&& F_2^D(x_N,Q^2,M) = \int\limits_{0}^{M_D/m} 
dy  
F_2^N \left( \frac{x_N}{y}, Q^2\right) f^{N/D}_M(y),
 \label{f2ms}
\end{eqnarray}
where ``the effective distribution'' of nucleons in the deuteron
 is defined by
  \begin{eqnarray}
&&\!\!\!\!\!\!\!\!\!\!\!\! f_M^{N/D}(y)  \label{fndm}\\
  &=&  \frac{1}{2M_D}\int \frac{d^3{\bfgr p}}{(2\pi)^3} 
\delta\left ( 
y - \frac{M_D-w+p_3}{m}
\right) \Theta(y)
\frac{1}{2\omega M_D^2 (M_D-2\omega  )^2}
\nonumber \\
&& 
{\sf Tr}\left\{
\bar\phi_M(p_0,{\bfgr p})
 (\hat p_1+m)(\gamma_0+\gamma_3) (\hat p_1+m)
 \phi_M(p_0,{\bfgr p}) (\hat p_2+m)
\right \}_{p_0 = \frac{M_D}{2}-\omega }.
\nonumber
\end{eqnarray}

Two sum rules can be written down
for the effective distribution $f^{N/D}_M(y)$:
\begin{eqnarray}
&&\int\limits_{0}^{M_D/m} f^{N/D}_M (y) dy= \langle D| \hat Q |D \rangle_M = 1,
 \label{ch0} \\
&&\int\limits_{0}^{M_D/m} y f^{N/D}_M (y) dy= \langle D| 
\left (\Theta_N \right)_\mu^\mu  |D \rangle_M = 1-\delta_N,
 \label{emt}
\end{eqnarray}
where  $\hat Q \propto \bar \psi(x)\gamma_0\psi(x)$ is the vector charge
 and 
 $ \left (\Theta_N\right)_\mu^\mu \propto i
\bar \psi(x)\gamma_\mu\partial^\mu\psi(x)$ is the trace of
energy-momentum tensor. 
Eq.~(\ref{ch0})  presents   
  the vector charge conservation generalized for the  
  deuteron states with different $M$.
In spite of  such clear physical interpretation, 
some time ago it was
 a subject of some controversy~\cite{fs2,land,km}.
Indeed, the  derivation of sum rule (\ref{ch0})
contains some subtle points and equivalence between it
and the expression for the charge
\begin{eqnarray}
\!\!\!\!\!\! \langle D| \hat Q |D \rangle_M =  \frac{1}{2M_D}  \int \frac{d^4p}{(2\pi)^4} 
 \frac{1}{3}\sum_M {\sf Tr}\left\{
\bar\Psi_M(p_0,{\bfgr p})\gamma_0 \Psi_M(p_0,{\bfgr p}) (\hat p_2-m)
\right \} 
 \label{ch}
\end{eqnarray}
   is a non-trivial fact, particularly,
    because of the presence of the $\Theta$-function
   in eq.~(\ref{fndm}). This $\Theta$-function provide a correct kinematics
   in  variable $x_N$  but cuts out a part 
   of the integration domain in $d^3{\bfgr p}$. 
   This cutting of the integration
   interval in the polar angle $\theta$ leads to non-zero  
   contribution of the  matrix element 
   containing $\gamma_3$, which is proportional to $cos\theta$. 
The   sum rule (\ref{emt}) for the first moment of $f_M^{N/D}$ is of a
different nature, it presents the nucleon contribution 
into the total momentum of the deuteron~\cite{bls,uk} and
$\delta_N$ is a part of the total momentum carried by the non-nucleon
component (mesons). An {\em important} property of sum rules (\ref{ch0}) and
(\ref{emt}) is that their r.h.s. do not depend upon the deuteron spin
orientation.

 The  SFs $F_{1,2}^D$ and $b_{1,2}^D$ are now 
calculated as follows:
\begin{eqnarray}
\left \{
\begin{array}{c}
 F_1^D(x_N,Q^2) \\
  b_1^D(x_N,Q^2)
 \end{array}\right \} =
\int\limits_{0}^{M_D/m}\frac{dy}{y}
\left \{
\begin{array}{c}
f^{N/D}(y) \\
\Delta f^{N/D}(y) 
 \end{array}\right \} 
 F_1^N \left( \frac{x_N}{y}, Q^2\right) ,\label{fb1} \\[3mm]
\left \{
\begin{array}{c}
 F_2^D(x_N,Q^2) \\
  b_2^D(x_N,Q^2)
 \end{array}\right \} =
\int\limits_{0}^{M_D/m} dy
\left \{
\begin{array}{c}
f^{N/D}(y) \\
\Delta f^{N/D}(y) 
 \end{array}\right \} 
 F_2^N \left( \frac{x_N}{y}, Q^2\right) ,
\label{fb2} 
\end{eqnarray}
where distributions $f^{N/D}$ and $\Delta f^{N/D}$ are given by
\begin{eqnarray}
 f^{N/D} (y) &=& \frac{1}{3}\sum_M f_M^{N/D}(y),
 \label{fnd} \\
\Delta f^{N/D} (y) &=& f^{N/D}_1 (y) - f^{N/D}_0(y).
 \label{dfnd}
\end{eqnarray}

The sum rules for 
$f^{N/D}(y)$  and $\Delta f^{N/D}(y)$ follow from sum rules for 
$f^{N/D}_M(y)$ and definitions (\ref{fnd})-(\ref{dfnd}):
\begin{eqnarray}
&&\int\limits_{0}^{M_D/m} f^{N/D} (y) dy  = 
\frac{1}{3}\sum_M \langle D| \hat Q |D \rangle_M = 1,
 \label{ch1}\\
&& \int\limits_{0}^{M_D/m} y f^{N/D} (y) dy =
\frac{1}{3}\sum_M \langle D|  
\left (\Theta_N \right)_\mu^\mu |D \rangle_M 
 = 1-\delta_{N},
\label{emt1}\\
&&\int\limits_{0}^{M_D/m}\Delta f^{N/D} (y) dy  = 
  \langle D| \hat Q |D \rangle_{M=1} - \langle D| \hat Q |D \rangle_{M=0}=0,
 \label{ch2}\\
&& \int\limits_{0}^{M_D/m} y \Delta f^{N/D} (y) dy =
  \langle D|  
\left (\Theta_N \right)_\mu^\mu |D \rangle_{M=1}
- \langle D|  
\left (\Theta_N \right)_\mu^\mu |D \rangle_{M=0}
 = 0.
\label{emt2}
\end{eqnarray}

Sum rules for the deuteron  SFs $b_1^D$ and $b_2^D$
are the immediate result of combining eqs.~(\ref{ch2})-(\ref{emt2})
and (\ref{fb1})-(\ref{fb2}):
\begin{eqnarray}
\int\limits_0^1 dx_D b_1^D(x_D) =0, \quad
\int\limits_0^1 dx_D b_2^D(x_D) =0,
 \label{srbd}
\end{eqnarray}
i.e. in a full agreement with the sum rules (\ref{sr1}) and (\ref{sr2}).

An explicit expression  for the distribution function  $f^{N/D}_M(y)$ (and therefore of $f^{N/D}(y)$ and 
 $\Delta f^{N/D}(y)$) in terms of the components of the 
 Bethe-Salpeter amplitude (eq.~(\ref{fndm}))
  is quite cumbersome and it will be presented elsewhere. 

\section{Nonrelativistic formulae}

The nonrelativistic expressions for  $f^{N/D}(y)$ and 
 $\Delta f^{N/D}(y)$ can be obtained by using an analogy
of the charge densities calculated within the Bethe-Salpeter 
formalism and corresponding densities calculated with wave
functions, i. e.:
\begin{eqnarray}
&&\!\!\!\!\!\!\!\!\!\!\frac{1}{2M_D}\int\limits_{-\infty}^{+\infty}\frac{ dp_0}{2\pi} {\sf Tr}\left\{
\bar\Psi_M(p_0,{\bfgr p})\gamma_0 \Psi_M(p_0,{\bfgr p}) (\hat p_2-m)
\right \} \quad  \sim  \Psi_M^\dagger({\bfgr p})\Psi_M({\bfgr p}) ,
 \label{an0} \\
 &&\!\!\!\!\!\!\!\!\!\!\frac{1}{2M_D}\int\limits_{-\infty}^{+\infty}\frac{ dp_0}{2\pi}  {\sf Tr}\left\{
\bar\Psi_M(p_0,{\bfgr p})\gamma_3 \Psi_M(p_0,{\bfgr p}) (\hat p_2-m)
\right \}
 \quad  \sim   \frac{p_3}{m}\Psi_M^\dagger({\bfgr p}) \Psi_M({\bfgr p}),
 \label{an3}
\end{eqnarray}
where $\Psi_M({\bfgr p})$ is a  nonrelativistic wave function of the deuteron
(do not confuse with the Bethe-Salpeter amplitude, $\Psi_M(p_0, {\bfgr p})$!)
The well-known result for the spin-independent distribution is immediately
 reproduced (see e.g.~\cite{msm,uk,bls}):
 \begin{eqnarray}
  f^{N/D}_{n.r.} (y) &=& \int \frac{d^3{\bfgr p}}{(2\pi)^3} 
  \delta\left (y-\frac{M_D-w+p_{3}}{m} \right ) 
\Theta \left (y \right ) 
\nonumber \\[2mm] &&\quad
(1+\frac{p_{13}}{m})
  \frac{1}{3}\sum_M \Psi_M^\dagger({\bfgr p})\Psi_M({\bfgr p})\nonumber \\
    &=&   \int \frac{d^3{\bfgr p}}{(2\pi)^3} 
  \delta\left (y-\frac{M_D-w+p_{3}}{m} \right ) 
\Theta \left (y \right ) 
\nonumber \\[2mm]&&  \quad
(1+\frac{|{\bfgr p}|cos\theta}{m})
  \left \{  u^2(|{\bfgr p}|) +
  w^2(|{\bfgr p}|) \right \},
 \label{fndnr}
\end{eqnarray}
where $u$ and $w$ are the $S-$ and $D$-wave components of the deuteron
wave function, respectively.
The presence of the $\Theta$-function in the r.h.s. of eq.~(\ref{fndnr})
slightly violates the sum rule (\ref{ch1}). However, this usually
phenomenologically is not noticeable, since the only region of large momenta,
$|{\bfgr p}| > 0.7$~GeV, is affected by the $\Theta$-function and it does not 
contribute much to the norm of the deuteron wave function. 

For distribution $\Delta f^{N/D}(y)$, one gets:
 \begin{eqnarray}
\Delta f^{N/D}_{n.r.} (y) &=&   \int \frac{d^3{\bfgr p}}{(2\pi)^3} 
  \delta\left (y-\frac{M_D-w +p_{ 3}}{m} \right ) 
\Theta \left (y \right ) 
\nonumber \\[2mm]
&& \quad (1+\frac{p_{13}}{m})
  \left \{   \Psi_1^\dagger({\bfgr p})\Psi_1 ({\bfgr p})
  - \Psi_0^\dagger({\bfgr p})\Psi_0({\bfgr p})\right \}\nonumber \\
  &=&   \int \frac{d^3{\bfgr p}}{(2\pi)^3} 
  \delta\left (y-\frac{M_D-w+p_{3}}{m} \right ) 
\Theta \left (y \right ) 
\nonumber \\[2mm]
&&\quad  (1+\frac{|{\bfgr p|}cos\theta}{m}) P_2(cos\theta)
 \frac{3}{2} w(|{\bfgr p}|)
  \left \{ 2\sqrt{2}u(|{\bfgr p}|) + w(|{\bfgr p}|) \right \},
 \label{dfndnr}
\end{eqnarray}
where $P_2$ is the Legendre polynomial.

Again, the  sum rule (\ref{ch2}) is
broken by the presence of the $\Theta$-function in (\ref{dfndnr}).
Neglecting it, one gets
 \begin{eqnarray}
\int\limits_0^1 dx_D b_1^D(x_D)&\propto &
\int\limits_0^{M_D/m} \Delta f^{N/D}_{n.r.} (y) dy\nonumber\\
& \propto&
 \int\limits_{-1}^{1} d(cos\theta) (1+\frac{|{\bfgr p|}cos\theta}{m}) P_2(cos\theta)
 =0,
 \label{sr1nr}
\end{eqnarray}
where the orthogonality property of the Legendre polynomials is used.

A deviation from zero, caused by the  $\Theta$-functions is not large compared
to 1,
but {\em anything} is large compare to 0! One can {\em artificially}
adjust formula (\ref{dfndnr}) to  fulfill this sum rule. For instance,
{\em small} corrections to the normalization of the both terms with
$M=1$ and $M=0$ can be made to satisfy the sum rule in  form
(\ref{ch2}). However, the situation with the second  sum rule,
(\ref{sr2}) and (\ref{srbd}), is more difficult and  can not be fixed by
any simple adjustments of the normalizations. 
Similar to  eq.~(\ref{sr1nr}), it can be written (neglecting the 
$\Theta$-function!):
 \begin{eqnarray}
&&\int\limits_0^1 dx_D b_2^D(x_D)\propto 
\int\limits_0^{M_D/m} y  \Delta f^{N/D}_{n.r.} (y) dy\nonumber\\
\quad
\quad & \propto&
 \int\limits_{-1}^{1} d(cos\theta) 
 {(M_D-w+|{\bfgr p|}cos\theta) }
 (1+\frac{|{\bfgr p|}cos\theta}{m}) P_2(cos\theta)
 \neq 0.
 \label{sr2nr}
\end{eqnarray}
Thus, there is no reason for this sum rule to be satisfied with
the nonrelativistic distribution function (\ref{dfndnr}).

In conclusion to this section, the nonrelativistic formulae, in principle,
violates the sum rules for the  SFs $b_{1,2}^D$. However, one can 
still hope that it will be a small effect, one not noticeable in practice.
The illustrative numerical  calculations are given in the next Section, 
this helps one
to understand a quantitative side of the effects.

\section{Numerical calculations}

In this Section results of the numerical calculations are presented.
The SFs of the deuteron $b_{1,2}^D(x)$ are calculated within both
the relativistic   and nonrelativistic approaches. 

The relativistic calculations are based on the formulae (\ref{fndm}), (\ref{fb1})-(\ref{dfnd})
and utilizes the realistic 
  Bethe-Salpeter amplitude
for the deuteron calculated in ref.~\cite{ukkk}. This amplitude is essentially just 
a different presentation of the amplitude obtained by Zuilhof and Tjon~\cite{tjond}.
An important feature of the calculations is a numerical ``inverse Wick rotation''
of the amplitude. This has been done by 
expanding 
a non-singular part of the matrix elements (\ref{fndm})
 into a series in $p_0/m$, up to the 
fourth order. 
Singularities in the nucleon propagators 
 have been treated exactly. Note, the numerical approximation made
  can potentially cause a violation of the exact sum rules.
  
The  nonrelativistic calculations, eq.~(\ref{dfndnr}), uses
the realistic wave function
of the deuteron in the Bonn potential~\cite{bonn}.
Another ingredient of the calculations, the nucleon  SFs 
$F_{1,2}^N(x,Q^2)$, is taken from ref.~\cite{amb} at $Q^2 = 10$~GeV$^2$.
 The results are neither
very sensitive to the particular choice of the parametrization for 
the nucleon  SFs nor to the $Q^2$-dependence of them.

The distribution functions $\Delta f^{N/D} (y)$  are calculated 
and the results are presented in Fig.~\ref{db2}. The relativistic (solid line)
and nonrelativistic (dotted line) give very similar behavior of the
distribution function. Indeed, it is difficult to distinguish between them,
not speaking about making  definite conclusions. The third line in the Fig.~\ref{db2}
is given for an illustration, and presents 
$y\Delta f^{N/D}(y)$  for the relativistic
calculations.
The calculation of the sum rules is more representative.
To understand the scale of effects, which are discussed below, 
it is customary to define auxiliary quantities: 
\begin{eqnarray}
\int\limits_{0}^{M_D/m}
 {\sf Abs}\left ( \Delta f^{N/D}(y)\right )dy  \simeq 
 \int\limits_{0}^{M_D/m}
 {\sf Abs}\left ( y \Delta f^{N/D}(y)\right )dy \simeq 0.14.
\label{aux}
\end{eqnarray}
The Bethe-Salpeter  
and   nonrelativistic Bonn calculations give the same result 
in (\ref{aux}), with accuracy of $\sim 5\%$. Thus, 
the effective distribution functions, $ \Delta f^{N/D}$, are an order 
of magnitude smaller than the usual spin-independent distributions $f^{N/D}$
normalized on 1. This is not a very important circumstance, but it works
against accuracy in the numerical calculations, since $ \Delta f^{N/D}$
is a difference of two functions normalized on 1 ($M=1$ and $M=0$). 
Numerically, the   sum rule (\ref{sr1}) 
 is satisfied both in relativistic and nonrelativistic calculations
 with good accuracy, despite the approximate
 numerical ``inverse Wick rotation'' and
   the discussion after eq.~(\ref{dfndnr}).  The correspondent integrals are
   $\sim 5\cdot 10^{-4}$   and $\sim 3\cdot10^{-5}$ and they should be compared 
   to the estimate~(\ref{aux}).
 The sum rule  (\ref{ch0})  may be used
to improve distributions  $\Delta f^{N/D}$ by making integrals for
$f^{N/D}_1$ and  $f^{N/D}_0$ {\em exactly } the same.
However, this does not lead to a
significant variation of results for  SFs, except $x\to 0$
for $b_1^D(x)$. 

The behavior of the $b_1^D(x)$ at $x\to 0$ deserves to be considered
more closely, especially for  numerical calculations, since
the nucleon function $F_1^N(x)$  can be divergent at small $x$.
Unfortunately it is impossible to estimate $b_1^D(0)$ exactly for
the realistic  SFs $F_1^N$. However, a contribution of
singularity can be evaluated. Indeed, let us assume a singular
behavior as $F_1^N\sim C/x$,
then eq.~(\ref{fb1}) leads for small $x$ to
\begin{eqnarray}
\!\!\!\!\!\!\!\! b_1^D (x\to 0) 
&\sim&  \frac{C}{x} \int\limits_{x}^{M_D/m} \Delta f^{N/D} (y)dy 
= \frac{C}{x} 
\left\{
\int\limits_{0}^{M_D/m} 
 -  \int\limits_{0}^{x} \right \}
 \Delta f^{N/D} (y) dy\nonumber \\
 &\sim& \frac{C\cdot Z}{x} - C \Delta f^{N/D} (0),
 \label{estd}
\end{eqnarray}
where $Z = 0$ in exact relativistic formula, but it
can be a small number in numerical 
calculations or in the nonrelativistic formalism. 
Thus, the limit of the deuteron  SF $b_1^D(x)$ as $x \to 0$ 
is a constant, but one has to exercise great care in performing 
the numerical
computations, since any error leads to
a divergent behavior  at small $x$. In this context, 
an adjustment of norms of the two terms in formulae (\ref{dfnd}) and
(\ref{dfndnr}) has a meaning of
subtraction of  the numerical error from $b_1^D$ at small $x$.

The situation with the second sum rule (\ref{sr2}) is quite different.
Numerically it is broken more significantly than the previous one.
Correspondent integrals are $\sim 1\cdot 10^{-3}$ and $\sim 3\cdot 10^{-3}$ 
for relativistic and nonrelativistic calculations respectively, i.e. about $0.7\%$
and $2\%$ compare to (\ref{aux}).
Therefore,  numerical approximations slightly damage the relativistic
formula. It is attributed to the numerical rotation to the
Minkowski space. An adjustment of the normalization, as it has been discussed,
slightly improves  the accuracy (to $0.5\%$). On the contrary, the result for
the nonrelativistic approach is stable with respect to any adjustments, since
it is defined by the formulae (\ref{sr2nr}). 

The  SFs $b_1^D$ and $b_2^D$ are calculated within 
two approaches as well. The results are shown in Fig.~\ref{b12} a) and b).
The behavior of the functions in Fig.~\ref{b12} a) suggests the validity of the 
sum rule (\ref{sr1}). At the same time,  the nonrelativistic
calculation for 
$b_2^D$ in  Fig.~\ref{b12} b) (dotted line) obviously
does not satisfy sum rule (\ref{sr2}). The main difference of
 the relativistic and nonrelativistic calculations is at small $x$, where
 these approaches give different signs for the  SFs.
 To illustrate the effect of the presence of the $\Theta$-function under integral
 in nonrelativistic formula (\ref{dfndnr}), the calculations have been done as well
 with a restricted
 interval of integration over  $|{\bfgr p}|$. The condition
 $|{\bfgr p}|< 0.7 $~GeV    corresponds to the ``softer'' deuteron wave function,
 but makes sum rule (\ref{sr2nr}) exact. 
 Correspondent  SFs are shown in Fig.~\ref{b12} a) and b) (dashed line).
 The result of this ``experiment'' is that the
 effect of $\Theta$-function is not quantitatively significant. 
 It also does not affect the principle conclusion about the second sum rule (\ref{sr2}),
 but makes the defect slightly smaller. This is understandable, since
 the sum rule breaking terms in (\ref{sr2nr}) is $\propto |{\bfgr p}|cos\theta$.

\section{Summary}
The  structure functions $b_{1,2}^D(x)$ of the deuteron 
have been studied within the covariant approach based 
on the Bethe-Salpeter formalism
for the deuteron. 
It is shown that the  nonrelativistic
convolution model  results in an incorrect
behavior of these structures at small $x$
and violates the exact sum rules. The importance of  
accurate relativistic calculations
for  $b_{1,2}^D(x)$ is demonstrated.

\section{Acknowledgements}
 It is a pleasure to acknowledge
 stimulating conversations with L. Kaptari, F. Khanna
  and O. Teryaev. I would like to thank D. White for reading the manuscript
  and comments.

\newpage

\centerline{\large \bf Figure captions}

\vskip 1cm 

 \begin{figure}[h]
\caption{Triangle diagram}
\label{tre}
\end{figure}

\begin{figure}[h]
\caption{$\Delta f^{N/D}(y)$. Curves: the BS - solid, the Bonn - dotted,
the BS multiplied on extra factor $y$ - dashed.}
\label{db2}
\end{figure}

 \begin{figure}[h]
\caption{$b_{1,2}^D(x)$. Curves: the BS - solid, the Bonn - dotted,
the   Bonn with cut - dashed.}
\label{b12}
\end{figure}

\phantom{.}

\newpage

 \let\picnaturalsize=N
\def\picsize{10cm}
\def\picfilename{tre-dis.ps}
\ifx\nopictures Y\else{\ifx\epsfloaded Y\else\input epsf \fi
\let\epsfloaded=Y
\centerline{\ifx\picnaturalsize N\epsfxsize
 \picsize\fi \epsfbox{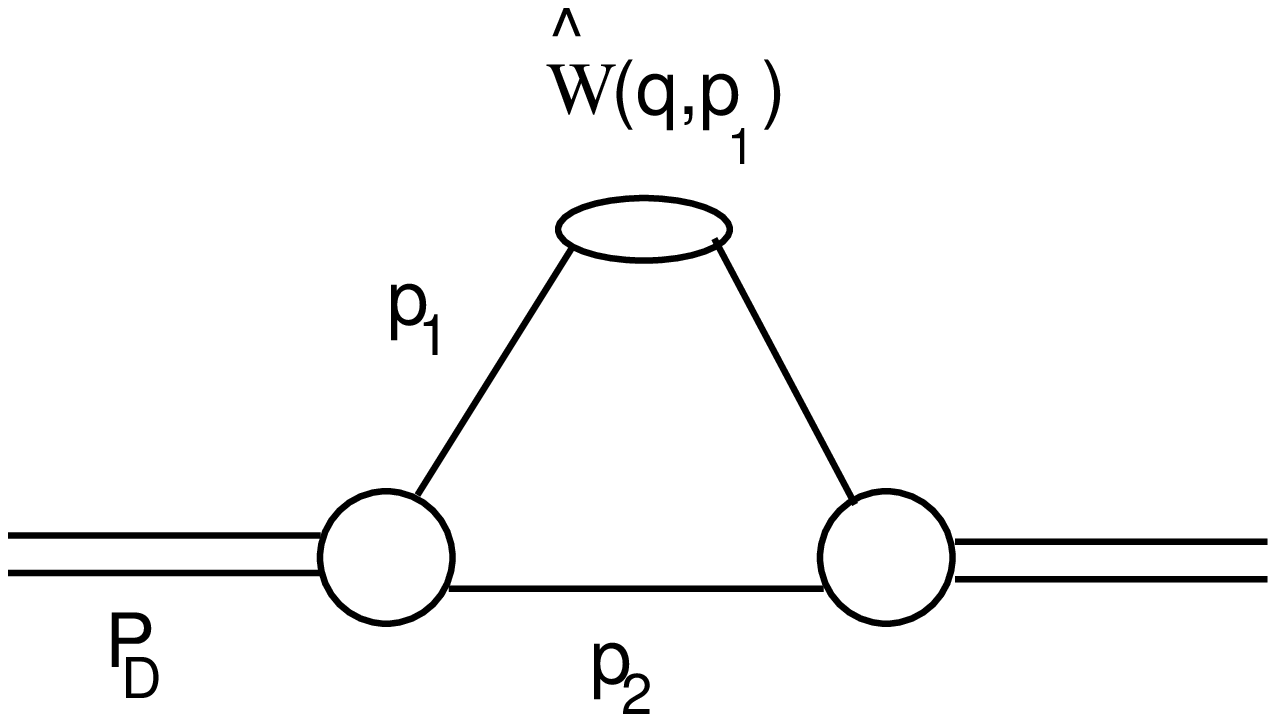}}}\fi
\vfill
Fig.~\ref{tre}.  A. Umnikov, Relativistic calculation...

\newpage

 \let\picnaturalsize=N
\def\picsize{15cm}
\def\picfilename{b2-dis.ps}
\ifx\nopictures Y\else{\ifx\epsfloaded Y\else\input epsf \fi
\let\epsfloaded=Y
\centerline{\ifx\picnaturalsize N\epsfxsize
 \picsize\fi \epsfbox{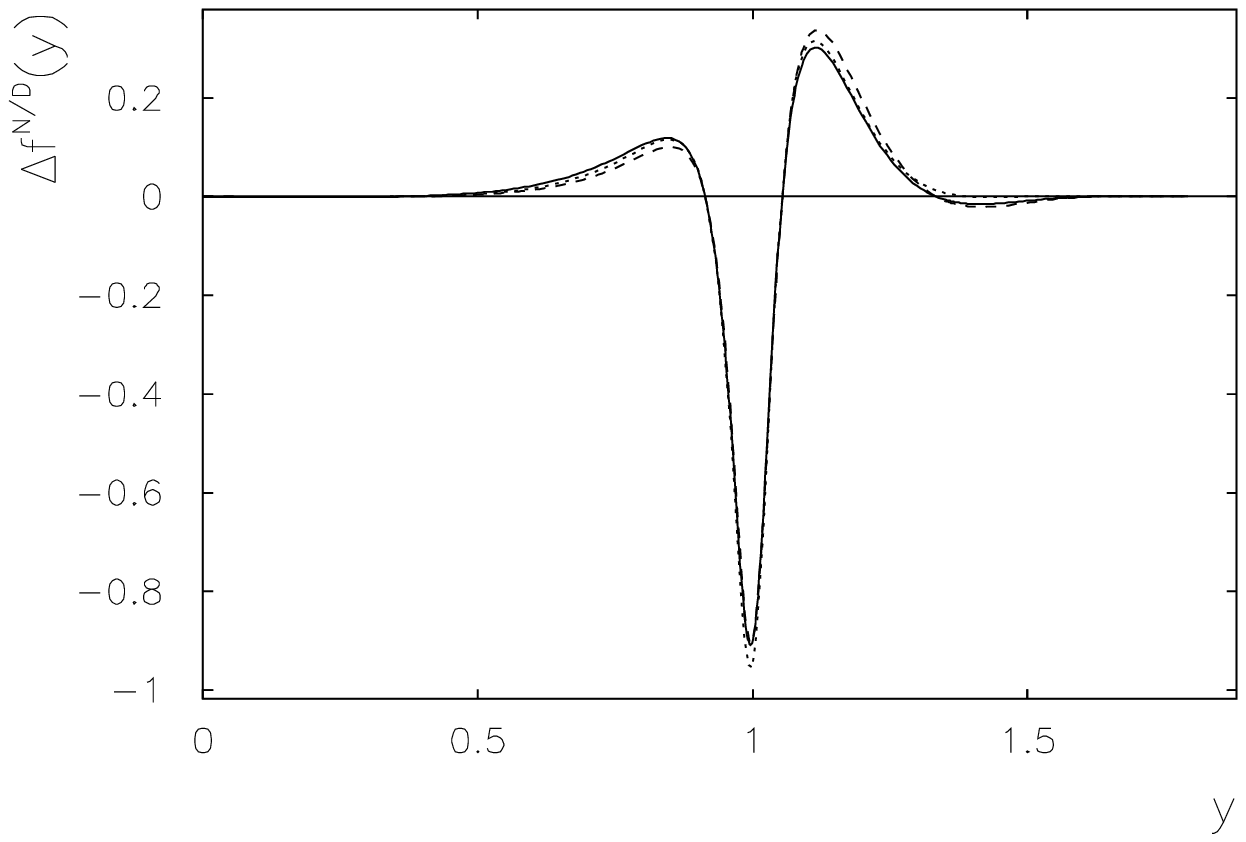}}}\fi
\vfill
Fig.~\ref{db2}. A. Umnikov, Relativistic calculation...

\newpage

 \let\picnaturalsize=N
\def\picsize{15cm}
\def\picfilename{b12.ps}
\ifx\nopictures Y\else{\ifx\epsfloaded Y\else\input epsf \fi
\let\epsfloaded=Y
\centerline{\ifx\picnaturalsize N\epsfxsize
 \picsize\fi \epsfbox{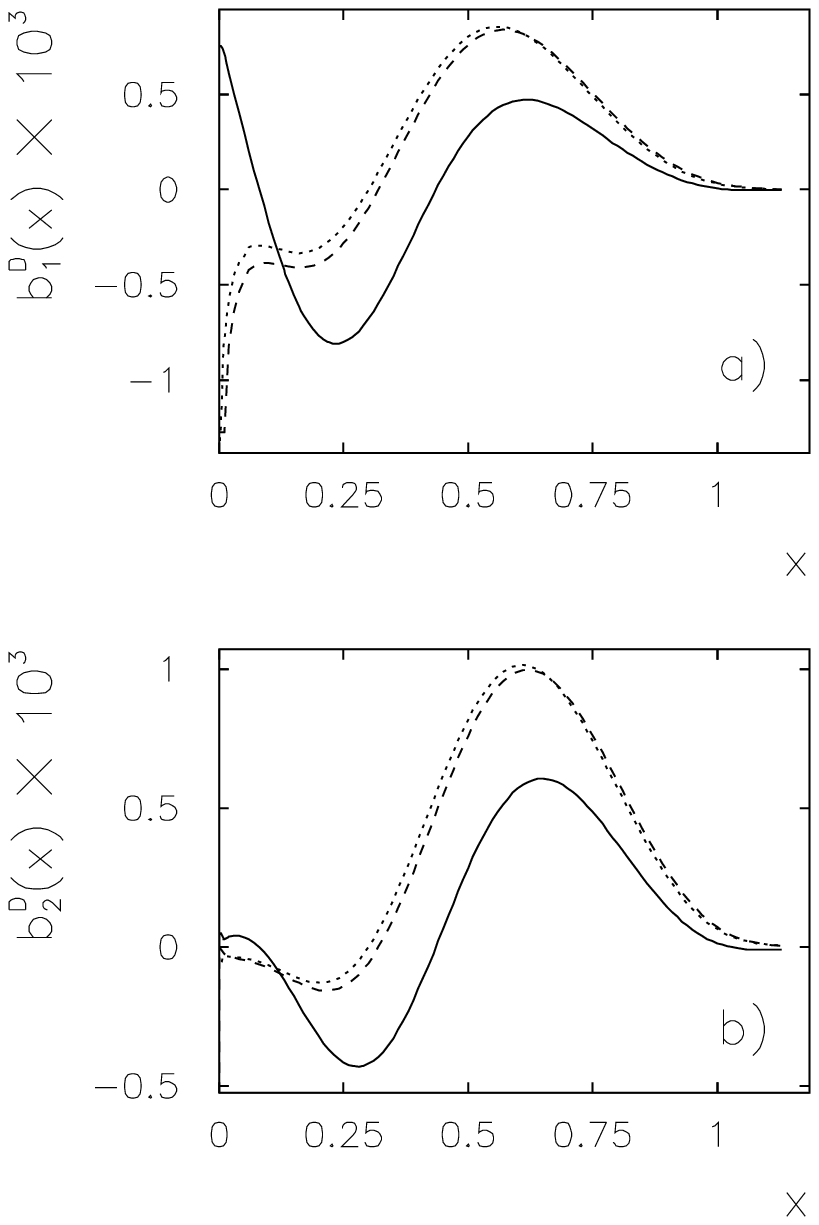}}}\fi

\vfill
Fig.~\ref{b12}.  A. Umnikov, Relativistic calculation...

 \end{document}